*Article*

# HiLTS©: Human in the Loop Therapeutic System: A Wireless-enabled Precision Medicine Platform for Brainwave Entrainment

Arfan Ghani *

Department of Computer Science and Engineering, School of Engineering and Computing, American University of Ras al Khaimah, United Arab Emirates

* Correspondence: Arfan.ghani@aurak.ac.ae; ghani786@gmail.com

**Abstract**

Epileptic seizures arise from abnormally synchronized neural activity and remain a major global health challenge, affecting more than 50 million people worldwide. Despite advances in pharmacological interventions, a significant proportion of patients continue to experience uncontrolled seizures, underscoring the need for alternative neuromodulation strategies. Rhythmic neural entrainment has recently emerged as a promising mechanism for disrupting pathological synchrony, but most existing systems rely on complex analogue electronics or high-power stimulation hardware. This study investigates a minimalist digital custom-designed chip that generates a stable 6 Hz oscillation capable of entraining epileptic seizure activity. Using a publicly available EEG seizure dataset, we extracted and averaged analogue seizure waveforms, digitised them to emulate neural front-ends, and directly interfaced the digitised signals with digital output recordings acquired from the chip using a Saleae Logic analyser. The chip's pulse train was resampled and low-pass-reconstructed to produce an analogue 6 Hz waveform, allowing direct comparison between seizure morphology, its digitized representation, and the entrained output. Frequency-domain and time-domain analyses demonstrate that the chip imposes a narrow-band 6 Hz rhythm that overrides the broadband spectral profile of seizure activity. These results provide a proof-of-concept for low-power digital custom-designed entrainment as a potential pathway toward simplified, wearable seizure-interruption devices for precision medicine and future healthcare devices.

## 1. Introduction

The semiconductor industry is entering a pivotal era, yet the global skills gap poses a major risk to its ambitions. In China, the "One Student One Chip" initiative empowers undergraduates to tape out actual processor designs, fostering hands-on hardware capability at scale [1]. In Europe, initiatives under the European Chips Act and the European Chips Skills Academy project highlight a projected shortfall of approximately 75,000 experts by 2030, specifically, 40,000 hardware engineers, 23,500 technicians and 11,300 software/data specialists [2]. Meanwhile, in the United States, the CHIPS and Science Act channels over US $52 billion toward reshoring semiconductor manufacturing, R&D and talent development; nonetheless, the workforce gap remains daunting, with estimates of 59,000-146,000 unfilled engineer and technician roles within the decade [3, 4]. In the Gulf region, the impetus is equally urgent: both the United Arab Emirates and Saudi Arabia have declared ambitions for AI-autonomous economies, yet talent surveys show that over 90 % of GCC organisations identify skills gaps as the foremost barrier to meaningful AI adoption [5, 6]. Without a home-grown semiconductor design and manufacturing ecosystem complemented with capable engineers, clean-room technicians and domain-specific research labs, the promise of regional autonomy in AI hardware cannot be realised.

This is where concerted capacity-building becomes essential. Hardware design houses, advanced training programmes and regional academic-industry collaborations are no longer optional; they are imperative. In this context, from our lab, we have taped out four microchips in 2025 with regional support and support from the IEEE Electron Devices Society. In this paper, we present the HiLTS platform, where several healthcare engineering designs were included as a system-on-chip neuromorphic test bed intended to catalyse novel AI hardware research. Such initiatives not only address the skills deficit but also anchor design-to-fabrication capability within the region, an indispensable step if the GCC is to move beyond chip consumption toward chip creation. The HiLTS platform includes a compact multiplixed system-on-chip platform that includes individual modules such as vagus nerve and multiphase backpain nerve stimulator design, seizure detection and entrainment core, Spiking Neural Network (SNN) based speech classifier, pacemaker, and a compact bare-metal PicoRV32 processor core to coordinate. However, for this paper, we are only considering the seizure detection and entrainment core. The discussion of other cores is beyond the scope of this article. An overview of the developed platform is shown below in Figure 1.

Rhythm is a universal phenomenon which can be observed in various natural environments such as physical, environmental and biological [7]. The human sensorimotor system possesses a remarkable ability to align itself with rhythmic patterns in the environment, from the steady pulse of light, the beat of music, to the circadian rhythms [8-10]. Research has shown that variations in rhythm, tempo, and auditory or visual stimuli can influence key physiological processes, including brainwave activity, heart rate, and motor coordination. This phenomenon, known as neural entrainment, reflects the brain's intrinsic tendency to synchronize with external rhythmic cues. Understanding this entrainment effect offers valuable insights into how rhythmic stimulation can modulate cognitive performance, emotional states, and motor behavior and forms the foundation for exploring digital, frequency-based interventions using neuromorphic chip-based systems [11-12].

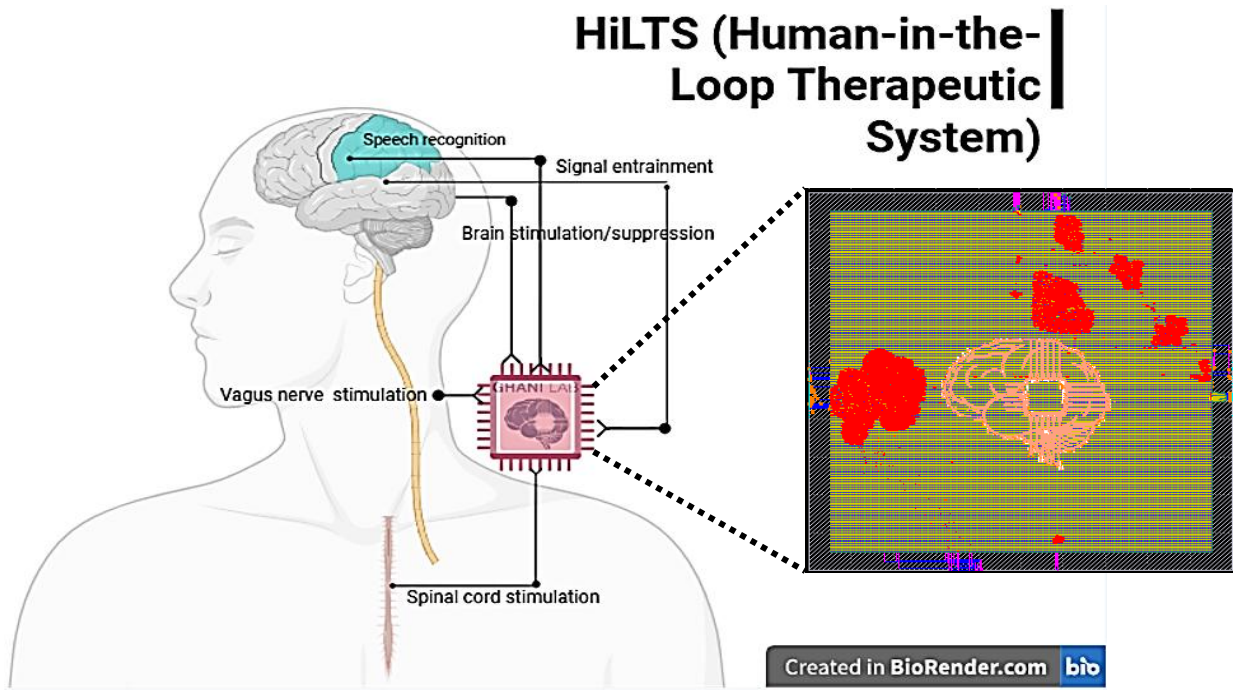

Figure 1: An overview of the HiLTS© platform

These insights highlight the importance of exploring how neural rhythms interact with and adapt to rhythmic patterns in our surroundings and cognitive processes. Epileptic seizures are a significant public health concern, affecting millions of individuals worldwide, including both adults and children. In 2022, epilepsy was identified by WHO as one of the top priorities and a special intersectoral global action plan on epilepsy and other neurological disorders for 2022-31 was adopted [13]. Epilepsy is characterized by highly synchronized abnormal neuronal activity in the brain due to a variety of underlying causes remains one of the most prevalent chronic neurological disorders worldwide, affecting approximately 50 million people [14]. The Global Burden of Disease database provides a comprehensive overview of the burden of disease in epilepsy of unknown cause, and therefore, this paper investigates and offers a potential solution to better understand and develop a solution from a technological point of view. To further underscore the importance, in the United Kingdom alone, current data indicate that roughly 6 – 10 people per 1,000 live with active epilepsy, and it remains one of the most common serious neurological disorders, with seizures posing risks of injury, psychological distress and social stigma. WHO data for the United Arab Emirates (UAE) and the wider Arab region are more limited, but regional reviews suggest a prevalence of active epilepsy in the Arabian Peninsula with a lifetime prevalence as high as ~12.9 per 1,000. These figures underscore the substantial global and regional burden of epilepsy, and highlight the critical need for effective management, treatment access and broader societal support [15-16]. Timely detection and intervention in the context of epilepsy are crucial for minimizing the risk of prolonged seizures and associated complications. The unpredictability of seizure onset poses a significant challenge for individuals living with epilepsy and their caregivers. Current treatment options, while effective for many, do not guarantee complete control over seizures, emphasizing the need for innovative approaches to monitoring and management. Advances in technology have paved the way for new methods of detecting and responding to seizure activity in real time, providing a pathway to improve patient outcomes.

This paper presents a potential technological solution aimed at addressing the urgent need for timely seizure detection and intervention through signal entrainment [17-19]. Brainwave entrainment method was used, which is a noninvasive technique that uses external signal patterns to guide the brain's activity. By exposing the brain to rhythmic stimuli at certain frequencies, it can synchronize its own electrical activity with the external rhythm. This alignment has the potential to influence both mental and physical states, helping to stabilize epileptic states, shape mood, focus, relaxation, or other physiological responses [20]. We introduce an open-source digital chip testbed to emulate brain activity for entraining signals immediately after the onset of seizures. Upon detection, the developed microchip could potentially be used to generate specific frequency rhythms intended to entrain and align seizure activity and restore normal brain function. To the best of the author's knowledge, this is the first of its type, which includes a hardware/software embedded platform for brain signal entrainment based on open-source design tools. The chip was designed with Skywater 130nm process and fabricated through Efabless by leveraging real-time data processing and responsive intervention. Our prototype aims to demonstrate the viability of such technological solutions to enhance the quality of life for individuals living with epilepsy, providing them and their caregivers with greater assurance and control over their condition. Through this innovative approach, we seek to contribute to the ongoing efforts to improve the management of epilepsy, ultimately aiming to reduce the burden of this challenging disorder on individuals, families, and healthcare systems worldwide. The specific contribution of this paper is as under:

1. an open-source, digitally implemented neuromodulation testbed for investigating and modulating epileptic seizures through rhythmic signal entrainment.
2. A custom-designed SkyWater 130 nm fabricated chip prototype capable of generating a stable 6 Hz oscillation for seizure entrainment.
3. a complete hardware signal analysis workflow, enabling direct comparison between extracted seizure signals, their digitised representations, and the chip-generated entrained outputs.
4. a system-level framework that lays the foundation for future closed-loop neuromodulation, including real-time EEG acquisition, automated seizure detection, and multi-frequency entrainment.
5. wireless IoT-based control using a mobile application, demonstrating the feasibility of portable and remotely managed neuromodulation

This paper is organised as follows:

Section 2 presents the software implementation and mathematical framework, followed by dataset preprocessing and the hardware simulation, implementation, and characterization. Section 3 briefly elaborates on the mobile application integration. Section 4 provides a detailed discussion, including limitations and future directions, and Section 5 concludes the paper with a summary.

## 2. Materials and Methods

*2.1. Software Design and Implementation*

Neural entrainment plays a critical role in coordinating oscillatory activity across brain regions. In epileptic seizures, abnormal neural synchronization often manifests as excessive, uncontrolled oscillatory coupling. By introducing controlled rhythmic stimulation, digitally generated precise frequencies, it may be possible to guide pathological brain rhythms back into stable, physiologically coherent states [21-23]. In this framework, a digital frequency-generation chip developed with open-source tools can deliver phase-controlled signals that interact with cortical oscillations, promoting adaptive entrainment rather than pathological suppression or synchronization. Such interventions could restore balance within neural networks, mitigating seizure onset and propagation through frequency-aligned entrainment of cortical rhythms [24]. To illustrate the principle, Figure 2 shows the high-level description of neural signal entrainment using digitally generated oscillatory waveforms through custom designed chip. The concept models how an external periodic signal, produced by a custom-designed frequency-generation chip, can influence and synchronize dysregulated brain activity. To illustrate this principle, the top plot in Figure 3 shows two independent oscillatory signals at 11 Hz (blue) and 13 Hz (red), representing neural populations operating at different intrinsic frequencies. These unsynchronized rhythms where oscillatory pools are not phase or frequency-aligned and therefore exhibit low coherence. In such a state, the peaks and troughs of the signals appear at different times, reflecting the chaotic nature of uncoordinated neural activity often observed in pathological conditions such as epilepsy [25-28]. Whereas the bottom plot shows two oscillatory signals with the same frequency (11 Hz) but a constant phase offset. This represents a phase-locked, entrained state in which the formerly independent oscillations have aligned under the influence of an external rhythmic drive. The constant phase difference indicates that the two signals are now coherently coupled, oscillating in a stable, synchronized relationship. In the neural context, this state models the successful entrainment of chaotic brain rhythms to an externally applied digital signal, restoring a balanced and regulated oscillatory pattern.

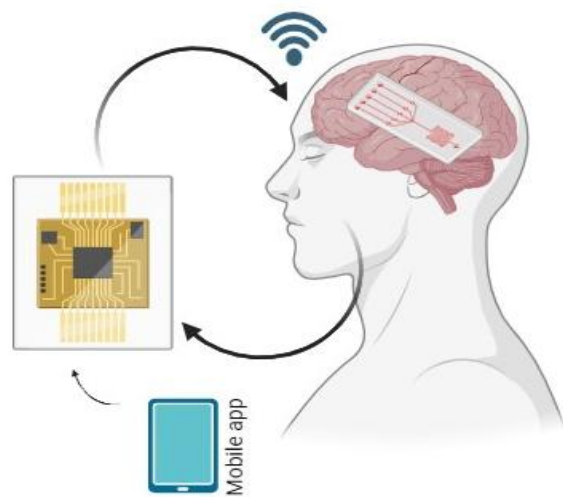

Figure 2. An embedded signal entrainment framework

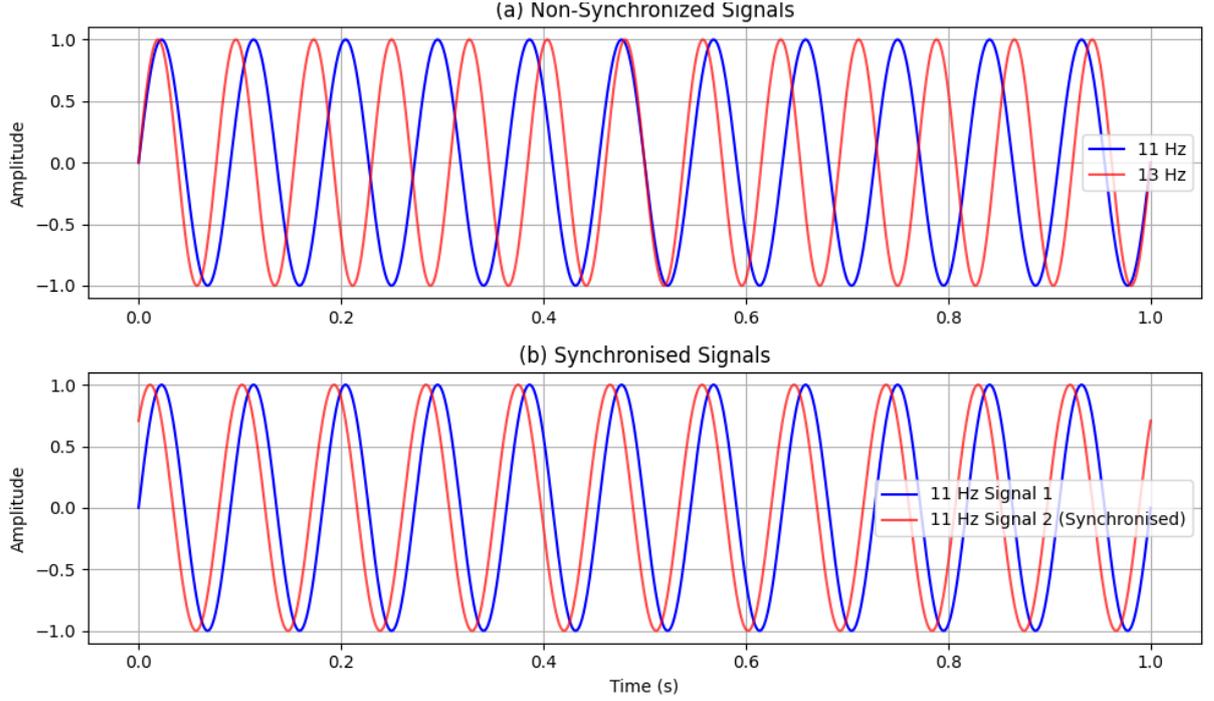

Figure 3. Illustration of neural entrainment using digitally generated oscillatory signals. (top) Independent oscillations at 11 Hz and 13 Hz represent dysregulated brain activity. (bottom) Phase-locked 11 Hz oscillations illustrate the entrained steady-state achieved when the brain's rhythm synchronizes with an external chip-generated signal.

Similar to the brain's electrical rhythms, signal entrainment involves the synchronization of an oscillatory system to an external periodic signal. In the context of brain waves, entrainment describes how an external rhythm generated by an external device could influence and synchronize with the brain's native rhythms, potentially bringing the brain wave activity closer to a desired frequency. In the context of this research paper where the brain's chaotic signal is not in sync with its expected normal rhythm, and hence classified as neural dysregulation. When a frequency with a specific rhythm generated by a custom-designed chip is introduced after the chaotic activity is detected, it works to entrain the chaotic signal by forcing it to align with the chip's generated frequency, thus bringing the brain signal back to a more regulated rhythm. The entrainment process can be modelled mathematically as a forced oscillator system[29]. The dysregulated signal, before the entrainment, is analogous to a free oscillator. Upon application of the pre-defined chip signal, the system is forced to adjust its phase and frequency to synchronize with the external stimulus. The chaotic signal can be modelled as random fluctuations with no inherent periodicity, and its frequency spectrum is spread across a wide range of values. The dysregulated signal could be represented by Equation 1.

$$x(t) = \eta(t) \qquad (1)$$

where $\eta(t)$ represents a random process. This signal fluctuates between positive and negative values without a defined frequency, making it irregular and unpredictable. Once the chaotic signal is detected, a trigger pulse is generated, and the external chip introduces a periodic forcing signal at a specific frequency, which could be modelled as Equation 2.

$$F(t) = A.pulse(t, f_{chip}) \qquad (2)$$

As shown in Eq.2, the pulse function represents a digital periodic pulse with frequency, $f_{chip}$, and $A$ as an amplitude, between 0 and 1, and $t$ is the time.

The entrainment of the dysregulated signal with the chip signal can be modelled using a coupled system of equations. As the neural chaotic signal is being forced by the periodic chip signal, a general form for the entrainment equation is given by Equation 3, and by substituting Equation 2, the final expression is shown in Equation 4.

$$\frac{d^2\theta}{dt^2} + 2\zeta\omega_0 \frac{d\theta}{dt} + \omega_0^2 \theta = F(t) \quad (3)$$

$$\frac{d^2\theta}{dt^2} + 2\zeta\omega_0 \frac{d\theta}{dt} + \omega_0^2 \theta = A.pulse(t, f_{chip}) \quad (4)$$

As shown in Equation 3, θ(t) is the phase of the signal, $\omega_0$ is the natural frequency of the brain signal prior to entrainment, ζ is the damping coefficient described as the internal resistance to entrainment, F(t) is the external forcing term, the chip signal, where $\omega_0$ and ζ are chosen based on the system's natural properties and the chaotic signal's behaviour. In order to achieve the signal entrainment, the phase synchronisation and locking happen when the phase θ(t) of the chaotic signal locks onto the phase of the periodic forcing chip signal, which corresponds to the system's response reaching a steady state, where the chaotic signal oscillates with the same frequency as the external signal. This phenomenon can be quantified by measuring the phase difference between the chaotic signal and the external signal. As the system entrains, the phase difference decreases, and both signals oscillate together with the same period.

*2.2. Dataset Pre-processing and Analysis*

The dataset used in this study consists of EEG recordings from 500 individual subjects, organized into five groups, each corresponding to a different EEG recording condition. [30]. Each subject's EEG data spans 23.6 seconds and is sampled at a rate of 173.3 Hz, resulting in 4097 data points per subject. These signals are divided into 23 chunks, each representing 1 second of data containing 178 data points. The EEG signals are stored in columns labelled X1 to X178, and the associated response variable, y, is represented in column 179. The y variable indicates the condition under which the EEG was recorded, with five labels ranging from 1 to 5. Label 1 corresponds to seizure activity, label 2 to the tumour region of the brain, label 3 to a healthy brain region, label 4 to eyes closed, and label 5 to eyes open. The goal of this dataset was to investigate and analyse the differences in EEG signals across these conditions, particularly the distinction between seizure and non-seizure activities. In the pre-processing step, the data is extracted from rows where the label '1' indicates seizure events. The dataset can be represented as a matrix X of size M×N, where M is the number of seizure events, and N is the number of time points. After extracting the dataset for seizure events, the plot for both superimposed and average signals is shown in Figure 4, where $X_i(t)$ could be represented as the EEG signal at time t for the $i^{th}$ seizure, and $I = 1, 2,…, M$ represents each seizure event, and $t=1,2,…, N$ represents the time points. In order to compute the average EEG signal across all seizure events, the mean of the signals was taken from all seizures at each time point. Mathematically, the average EEG signal $S_{avg}(t)$ at time *t* is given by Equation 5.

$$S_{avg}(t) = \frac{1}{M}\sum_{i=1}^{M} X_i(t) \quad (5)$$

In Equation 5, M is the total number of seizure events, and $X_i(t)$ is the EEG signal at time t for the $i^{th}$ seizure. The averaging process smooths out individual fluctuations and reveals the general trend of brain activity during a seizure, highlighting common features across different seizure events. It is particularly useful in identifying consistent patterns in the brain's electrical activity during seizures.

To analyze seizure signals, all the EEG segments that were labelled as seizure activity were extracted from the dataset. Each segment contains 178 samples of the EEG activity in microvolt-scale units, as shown in Figure 4. For the purposes of assessing the variability across seizure events and to determine any shared structural features, all waveforms for the seizure were plotted on the same axes, yielding a superimposition of the signals. This graphical representation highlights the diversity of seizure morphologies, reflected in differences in amplitude, shape, and timing, as well as recurring patterns that appear as darker regions where many signals overlap. In the middle plot, an average seizure waveform was computed by taking the mean across all seizure segments at each sample point. This yields a smooth, representative waveform that captures the dominant temporal structure commonly exhibited during seizure activity in this dataset.

To complement these qualitative observations, Shannon entropy was calculated for each seizure segment as a numerical measure of signal irregularity and complexity. Entropy was chosen because seizure EEG can vary widely in structure: some events exhibit highly rhythmic spikewave discharges, while others are more chaotic or irregular. Entropy, therefore, provides an objective way to quantify this variability. The amplitude values were first normalised into a probability distribution for each seizure segment using a normalised histogram. Then, the Shannon entropy of this distribution was calculated using Equation 6.

$$H = -\sum_{i=1}^{n} p_i \log(p_i) \quad (6)$$

where $p_i$ represents the probability associated with the ith amplitude bin. Segments with high entropy correspond to more irregular and less predictable seizure waveforms, whereas low entropy values indicate more structured, stereotyped activity. Plotting entropy across all seizure events reveals how the complexity of seizure dynamics varies from one event to another and provides a quantitative interpretation of the variability observed in the superimposed waveforms.

Taken together, the superimposed seizure plots, the average waveform, and the entropy distribution offer a comprehensive view of both the common underlying structure of seizure EEG activity and the substantial intrinsic variability that exists across individual seizure occurrences.

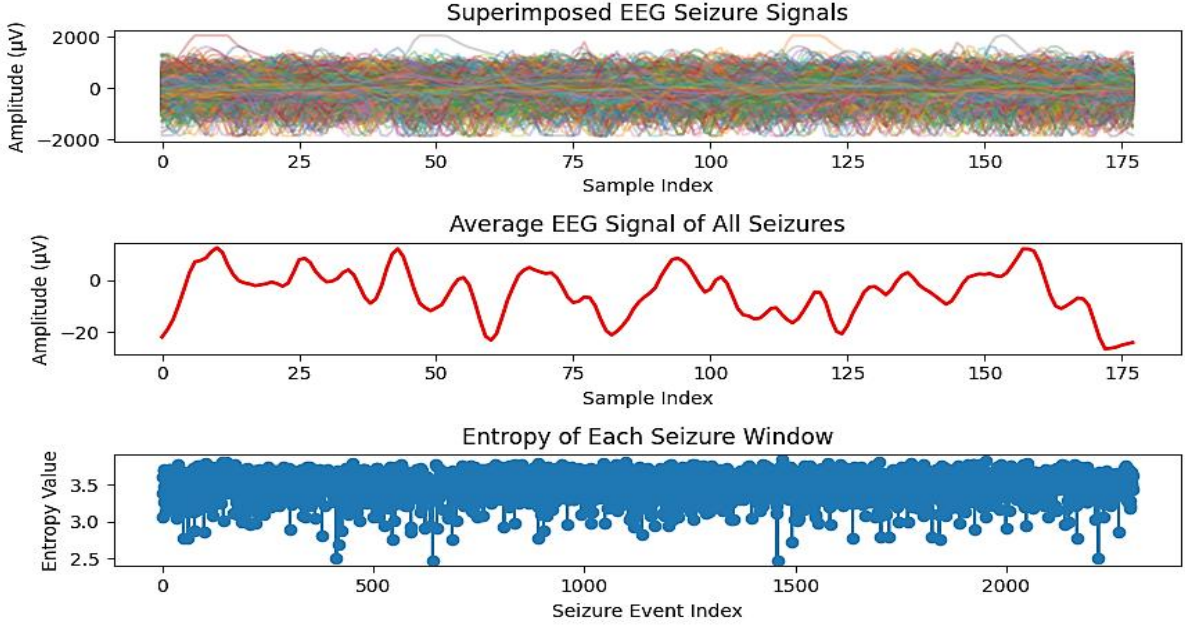

Figure 4: Superimposed seizure EEG signals, average waveform, and entropy distribution.

As EEG signals include noise hence to isolate the relevant brainwave frequencies, a bandpass filter was applied. The filter selectively allowed signals within a specific frequency range to pass while attenuating signals outside this range. For this study, the frequency range from 0.5 Hz to 40.0 Hz was selected, corresponding to the typical physiological brain rhythms delta, theta, alpha, beta and gamma. A Butterworth filter was used to select the desired frequency components, maximising its flat frequency response in the passband, while minimising distortion in the filtered signal.

The mathematical representation of a Butterworth filter can be represented by Equation 7.

$$H(f) = \frac{1}{\sqrt{1+(f/fc)^{2n}}} \qquad (7)$$

As shown in Equation 7, H(f) is the frequency response, f is the frequency, fc is the cutoff frequency, and n is the order of the filter. The filter's transfer function allows frequencies between the lower and upper cutoff frequencies to pass through while attenuating others. After filtering the EEG signals, the power was analysed in specific frequency bands. The FFT was applied to convert the signal from the time domain to the frequency domain, as shown in expression 8, where the FFT of a signal x(t) is defined as follows.

$$X(f) = \int_{-\infty}^{\infty} x(t)e^{-j2\pi ft}dt \qquad (8)$$

Where X(f) is the frequency-domain representation of the signal x(t). The squared magnitude of the FFT gives the power spectrum of the signal, as shown in Equation 9.

$$Power\ (f) = |X(f)^2| \qquad (9)$$

To calculate the band power by summing the squared magnitudes of the FFT coefficients within each frequency band of interest. The power within a specific frequency band is computed as equation 10, where $P_{band}$ is the band power for the frequency range.

$$P_{band} = \Sigma_{f_{low} \leq f \leq f_{high}} Power(f) \qquad (10)$$

The visualisation of the EEG signals and band power is shown in Figure 5.

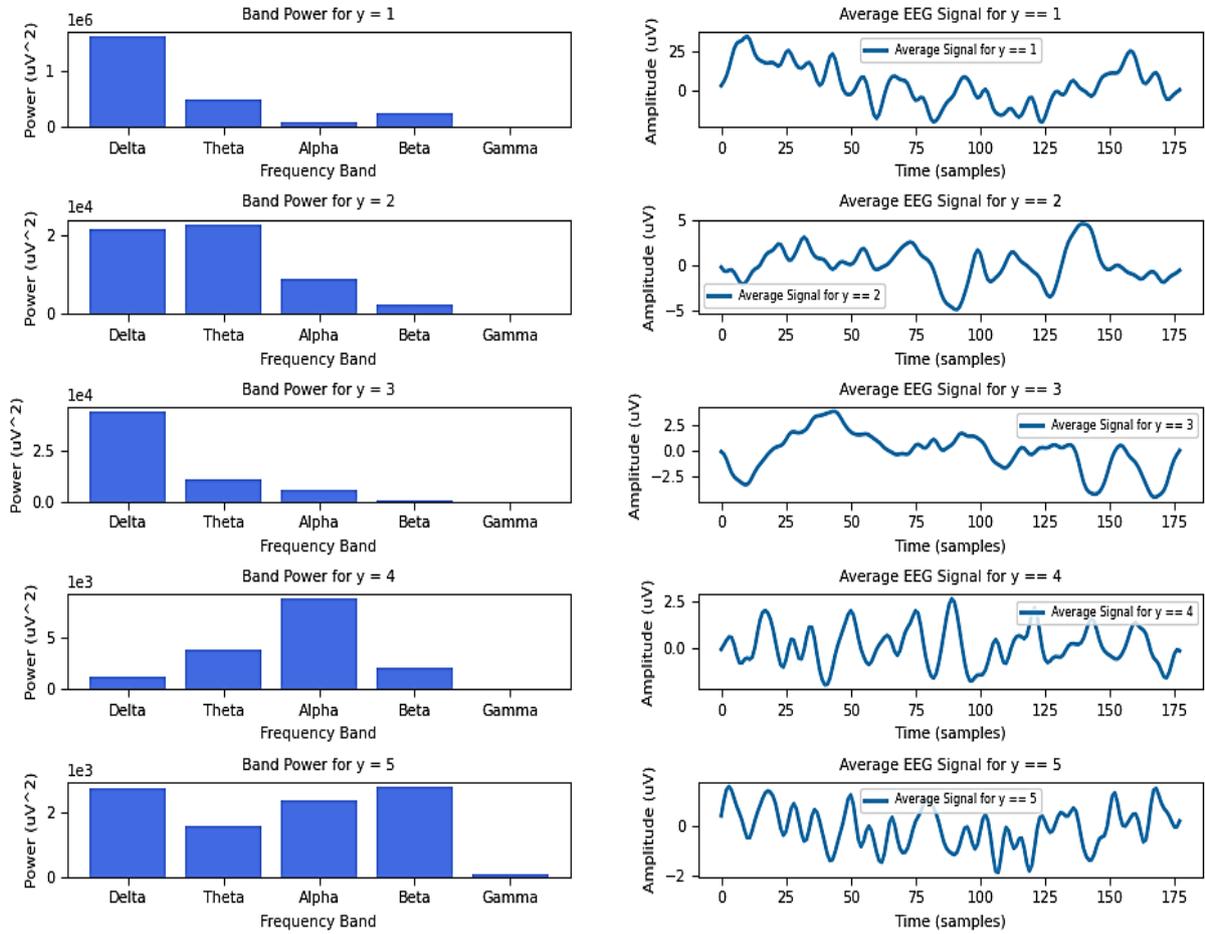

Figure 5: Visualisation of the EEG signals and corresponding band power

As shown in Figure 5, the bar plots present the power in the different frequency bands for each class (y = 1 to 5). The x-axis represents the frequency bands, and the y-axis represents the power in microvolts squared (µV²), which is the typical unit for EEG power. The right-hand side plot shows the average EEG signal for each class. The x-axis represents time (in samples), and the y-axis represents the amplitude of the EEG signal in microvolts (µV).

The analysis done on the EEG dataset provides important insights into the characteristics of the signals across the five clinical conditions represented by the labels y=1 to y=5. Separating the data according to class and then filtering it with a physiologically meaningful bandpass filter, 0.5 - 40 Hz, isolates the dominant brain rhythms, delta, theta, alpha, beta, and low gamma, removing high-frequency noise and slow drift often seen in raw EEG recordings. Computing the average waveform for each class provides a visual representation of the typical shape of the signals in time and reduces subject-specific variability, emphasizing characteristic signatures of seizure versus non-seizure activity. Running in parallel, the spectral analysis performed through the FFT and band-power computation quantifies the distribution of energy across the major EEG frequency bands.

Seizure activity is not defined by high-frequency content alone but rather by abnormal synchronisation, amplitude changes, and shifts in spectral energy. These distinctions are evident in the band-power bar plots: class 1 (seizure) signals show elevated power in the delta–theta range and broader spectral spread, consistent with paroxysmal discharges and hypersynchrony commonly seen during pre- and post-ictal events. In contrast, the non-seizure classes display more stable and band-specific power distributions. Hence, this combination of temporal averaging and frequency-domain analysis establishes that this dataset captures meaningful physiological differences between classes and further confirms that the extracted class-1 signals reflect seizure patterns even when the dominant frequency falls within a traditionally 'normal' frequency band. This establishes a reliable baseline against which the biological seizure dynamics will be compared to the behaviour of the hardware rhythm-generation chip used later for signal entrainment.

Before evaluating the hardware-level entrainment behaviour of the proposed custom-designed chip, it was necessary to construct a benchmark signal that correctly represents the characteristic dynamics of seizure activity. Each raw EEG trace contains 4097 samples over a duration of 23.6 seconds, yielding an effective sampling frequency of approximately 173.3 Hz. The average seizure waveform was computed by averaging all Class-1 segments (X1–X178), resulting in an analogue reference signal representing the dominant seizure morphology. Three transformations of this benchmark signal were then generated as shown in Fig. 6. The top plot shows the analogue seizure waveform derived from the dataset. The middle plot presents a digitised version of this signal, produced by applying a mean-based threshold which mimics the comparator circuits used within digital custom-designed hardware, as shown below.

$$D(n) = \begin{cases} 1 & if\ x(n) > \mu \\ 0 & otherwise \end{cases}$$

This illustrates how the biological signal would appear once converted into the binary event domain used by the chip. The third plot shows the logarithmic power spectrum of the seizure waveform, demonstrating its dominant energy distribution across low-frequency bands (4–8 Hz), consistent with rhythmic seizure discharges reported in clinical EEG literature. This frequency-domain characterisation is essential because signal entrainment in the hardware is fundamentally a frequency-selection and frequency-stabilisation problem.

These three representations, analogue, binary, and spectral, form the software benchmark against which the behaviour of the neural chip is evaluated. The subsequent simulations in the hardware section of the paper demonstrate how a chaotic seizure signal can be stabilised through an externally generated trigger pulse, after which the system entrains to a target healthy rhythm. Together, this validates the system-level concept: seizure activity can be digitally suppressed and replaced with a stable, healthy oscillatory pattern through hardware-based frequency entrainment. A flowchart of the system-level implementation is shown in Figure 7.

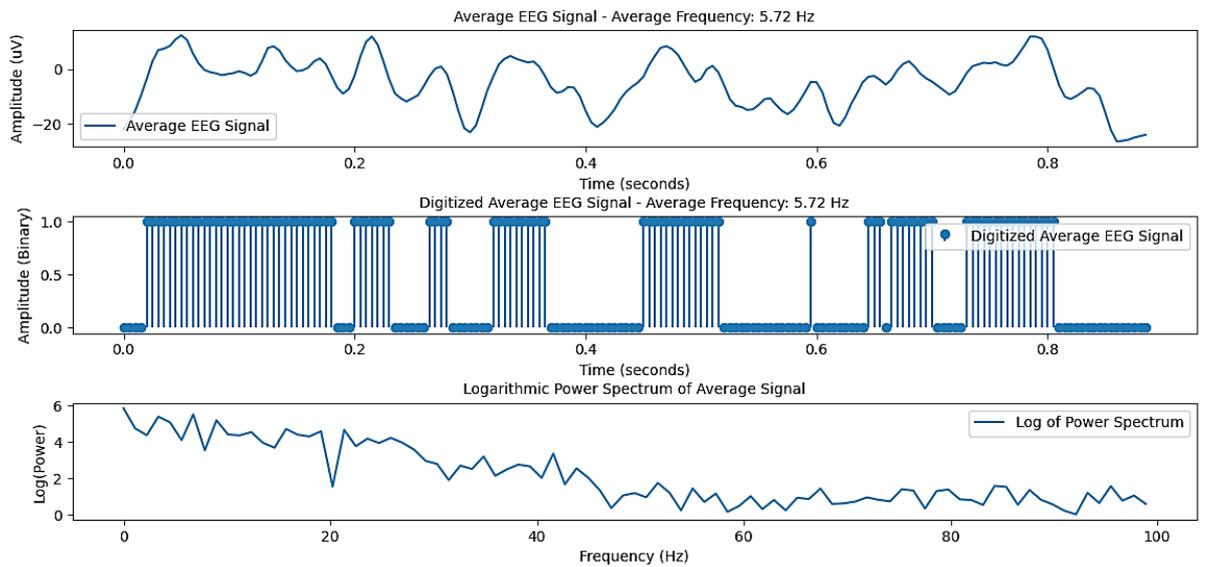

Figure 6: The top subplot shows the averaged analogue EEG signal (μV) over time (seconds). The middle subplot displays the digitised version of this signal, obtained by applying a threshold equal to the mean amplitude of the average signal, resulting in a binary representation. The bottom subplot presents the logarithmic power spectrum of the averaged signal, highlighting the distribution of signal power across frequencies (Hz) and emphasising dominant frequency components during seizure activity.

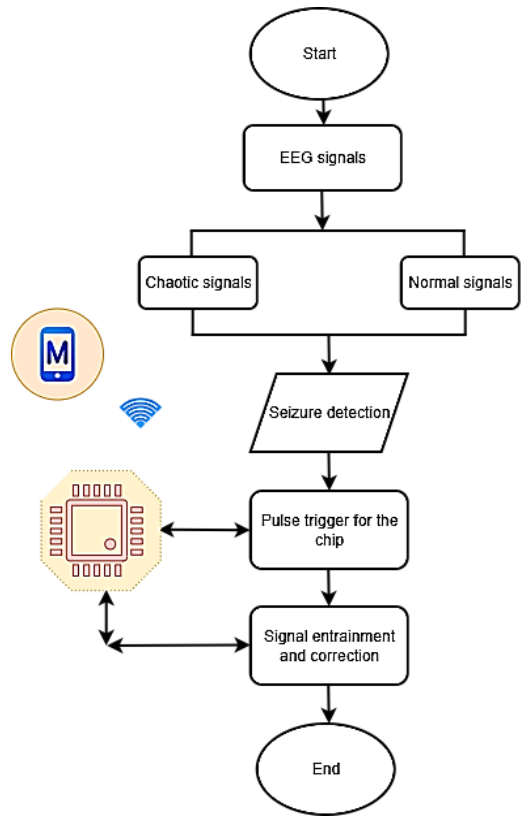

Figure 7: A system-level design flow for brainwave entrainment

*2.3. Hardware Design and Implementation*

To illustrate the principle of signal entrainment, a pseudo-chaotic signal scenario has been created for hardware emulation, as depicted in Figure 8. The top subplot illustrates the normal brain rhythm at 6 Hz, indicative of healthy neural oscillations, digitized into a binary pulse. The chaotic second subplot is created from high-frequency random noise to emulate the irregular and unpredictable activity found during epileptic seizures. The third subplot presents the trigger pulse, indicative of the detection of abnormal activity by the seizure detection device, which initiates the entrainment process.

After the trigger pulse, the digital chip generates a controlled 6 Hz rhythm. The recovered signal post-entrainment, shown in the bottom subplot, transitions smoothly from pseudo-chaotic activity to the stable normal rhythm. The proposed paradigm validates the signal detection and processing logic but also strengthens the scientific principle of chaotic-to-ordered neural entrainment. By emulating both irregular seizure-like activity and controlled rhythmic stimulation, the system shows how focused intervention can be used to reinstate stable brain oscillations principle underpinning the subsequent hardware implementation by means of open-source synthesis tools. The digital representation of pulses allows for the use of digital logic to create a real-world, real-time neurostimulation experiment.

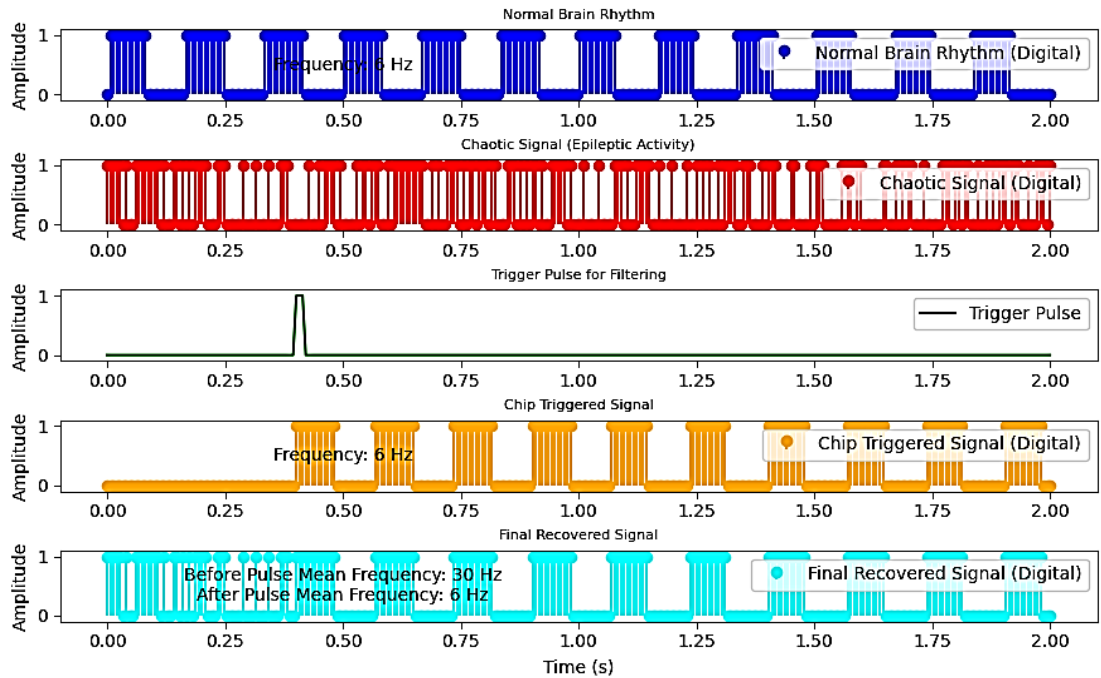

Figure 8: An illustration of the irregular brain signal entrainment with custom custom-designed chip

The simulation outlined here serves a critical role in the design and validation of an open-source hardware platform that processes real-time signals, especially in contexts of neurotechnology. In a real-world scenario, a device would be monitoring brain activity continuously. When the system detects seizure activity, it activates the trigger signal, causing the seizure detection paradigm to process the incoming signals. An open-source hardware synthesis tool, Yosys, was used to implement the design in hardware [31]. It serves as the foundation for synthesizing the design into a physical hardware implementation. Yosys is an open-source synthesis tool that takes high-level Verilog designs and translates them into a gate-level netlist that can be used to generate a hardware circuit. By providing this Verilog code in the Yosys format, the system could be implemented on hardware platforms where the chaotic signal detection and synchronisation could be applied in real-time. Therefore, the Yosys code is critical for bridging the gap between simulation and physical implementation, ensuring that the design can be validated in both a simulated environment and on actual hardware.

The hardware design synthesis activity diagram is shown in Figure 9, where the trigger pulse is controlled externally through the trigger_pulse input, and when this signal is asserted, the system switches from chaotic to a stable periodic signal that approximates a healthy brain rhythm. The chaotic signal is continuously updated based on random noise. When the signal exceeds a certain threshold, which is set to 8'b01111111 in the simulations, the system outputs a pulse, signalling that the chaotic activity has reached a level where intervention is needed. This pulse, once detected, sets the normal signal, normal_signal to a predefined value that represents the normal brain activity, and the chaotic signal is cleared. The testbench created provides a stimulus to the design by generating a clock signal and controlling the reset and trigger pulse inputs. The testbench simulates the behaviour of the system, applying a reset at the beginning and then allowing the chaotic signal to evolve. After a short period, the trigger pulse is asserted, causing the system to transition to the normal brain rhythm. The detailed output waveform activity is shown in Figure 9.

For this open-source emulation platform, the activity diagram demonstrates how a system can detect chaotic signals and synchronize to a desired normal rhythm. The chaotic signal represents an epileptic or irregular brain state, and the normal rhythm represents the target brain activity. The triggering mechanism simulates a medical intervention, such as a neural implant or brain stimulation device, that can restore the brain's activity to a stable, predictable state, analogous to a healthy 6 Hz rhythm. To further validate the paradigm, utilizing an open-source simulation environment such as Icarus Verilog, we can prototype and test signal correction mechanisms that could be deployed in embedded devices for brainwave monitoring and seizure detection. The open-source nature of the tool allows for flexibility in experimentation and optimization, making it an ideal choice for prototyping before moving to more complex hardware implementations. The goal is to develop systems that can detect irregular brain activity and correct it in real time, helping to manage neurological disorders. The RTL synthesis output waveform is shown in Figure 10.

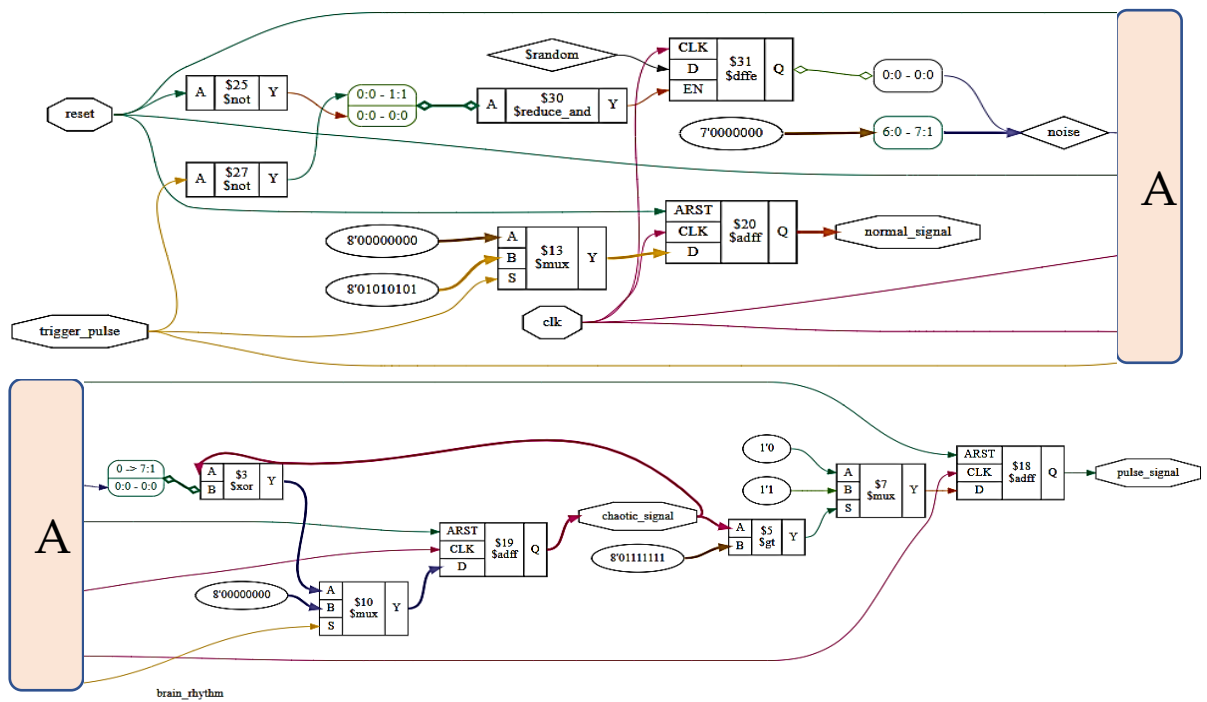

Figure 9: RTL synthesis activity diagram from chaotic signal detection to entrainment and back to normal rhythm

The main output from Yosys is the synthesized netlist, which details how the different components of the design are interconnected. This netlist contains information about the logic gates and their connections, showing how the chaotic signal generation, pulse detection, and filtering mechanisms work together. The netlist allows designers to verify that the intended logic and functionality have been preserved through the synthesis process. In the context of the provided code for the brain rhythm generator, the symbols in the Yosys output can be directly related to the various components defined in the design. The rectangles in the output represent the combinational logic used to generate the chaotic signal. Each rectangle corresponds to a specific logical operation defined in the code, helping to visualise how these operations combine to produce the chaotic behaviour.

The D flip-flops (DFFs) shown in the output represent the storage elements used to hold values at each clock cycle. For instance, the chaotic signal and restored signal are all stored in D flip-flops. These components capture the state of their respective signals on the rising edge of the clock, ensuring that the circuit behaves synchronously. Rhombuses, which signify decision points, relate to the logic that determines when the pulse signal is activated. The pulse signal logic checks if the chaotic signal exceeds a certain threshold. This decision-making process is crucial for detecting chaotic behaviour, and the corresponding rhombus in the Yosys output visually represents this conditional logic. The lines connecting these symbols illustrate the flow of signals, such as from the chaotic signal to the pulse signal logic. These connections are critical for understanding how data flows through the circuit and how the different components interact with each other. Overall, the Yosys outputs facilitate a comprehensive understanding of the design, enabling designers to verify correctness, optimize performance, and prepare the design for further stages such as implementation on an ASIC.

To emulate actual hardware behavior, hardware simulations were performed in System Verilog using a 100 MHz clock, where reset initializes all signals, and the simulation tracks the evolution of chaotic, filtered, and restored signals over time, as shown in Figure 10. This setup shows chaotic_signal fluctuating unpredictably, low_pass_filter smoothing these fluctuations, when the chaotic signal crosses the threshold, and restored_signal aligning with the chip_signal, illustrating how chaotic dynamics can be guided by a regular reference through strong entrainment.

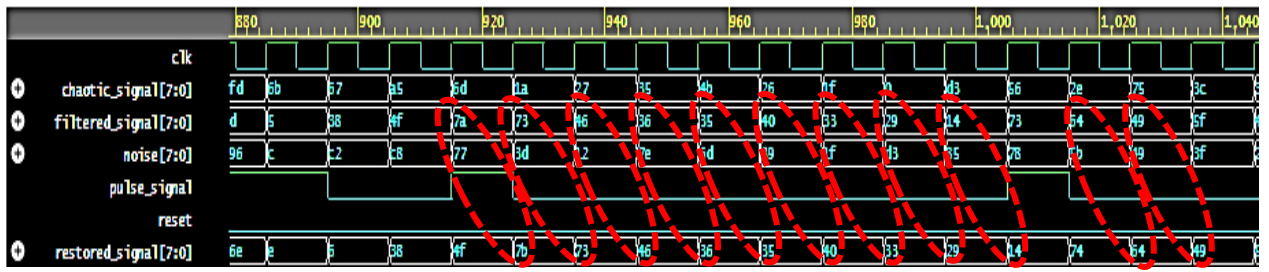

Figure 10: Hardware simulation of signal entrainment

The waveform generated by the simulation in Figure 10 provides a complementary view of the system's behaviour over time, illustrating how the input signals and output signals evolve during the simulation. The waveform allows for the observation of signal transitions, which is crucial for verifying that the FSM is triggering the expected actions and that the output signals correspond to the correct state at any given moment. The waveform can also provide insights into the timing of various signals, helping to ensure that all transitions occur in the correct sequence and at the right times, which is particularly important in real-time signal processing applications where precise timing can be a matter of critical importance.

*2.4. Chip Integration and Verification*

Following the Yosys-based hardware simulations, the custom chip was fabricated through the eFabless platform using the SkyWater 130 nm process, which provides a low-power, cost-effective fabrication flow using open-source PDKs [32]. The fabricated chip implements the chaotic-to-ordered brain rhythm entrainment logic, which was previously validated in software simulations. Verification was performed at multiple levels. Post-fabrication testing involved monitoring the chip output using a Saleae logic analyzer [33] and an oscilloscope to ensure correct frequency generation and trigger pulse timing. Functional verification was carried out by applying digital inputs simulating epileptic seizure events and comparing the chip output to the software simulation benchmarks. The chip successfully produced controlled 6 Hz rhythmic signals following detection of a pseudo-chaotic input, demonstrating correct signal entrainment behavior. Additionally, verification confirmed that the transition from chaotic to ordered signals adhered to the expected waveform profiles, maintaining amplitude, frequency, and timing fidelity. These tests validated that the chip can serve as a practical hardware platform for neuro-modulatory applications, capable of stabilizing irregular neural activity through controlled signal generation. The hardware characterization setup is shown in Figure 11, and chip-generated signals are shown in Figure 12.

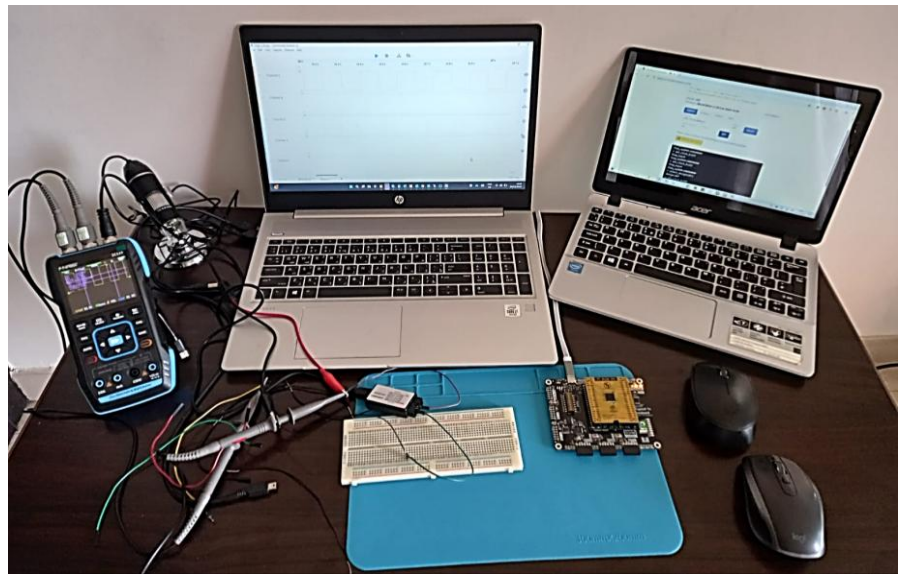

Figure 11: Chip characterization setup

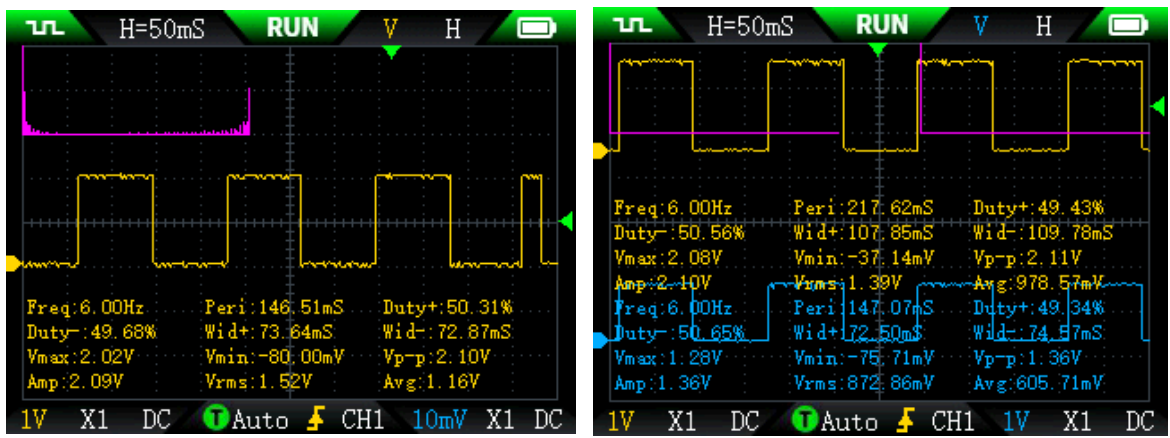

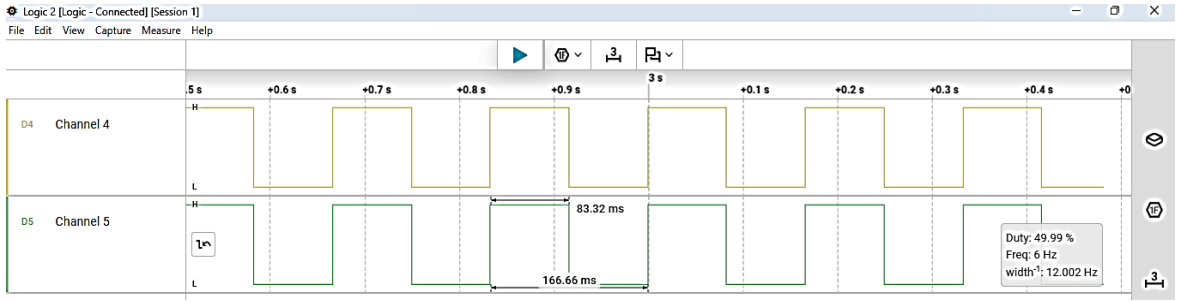

Figure 12: Chip-generated control entrainment signals top (oscilloscope), bottom (Saleae logic analyzer)

To evaluate the neural chip's interaction with the seizure signal, digital output from the chip was collected using a Saleae Logic Analyser. The chip output is a digital square-

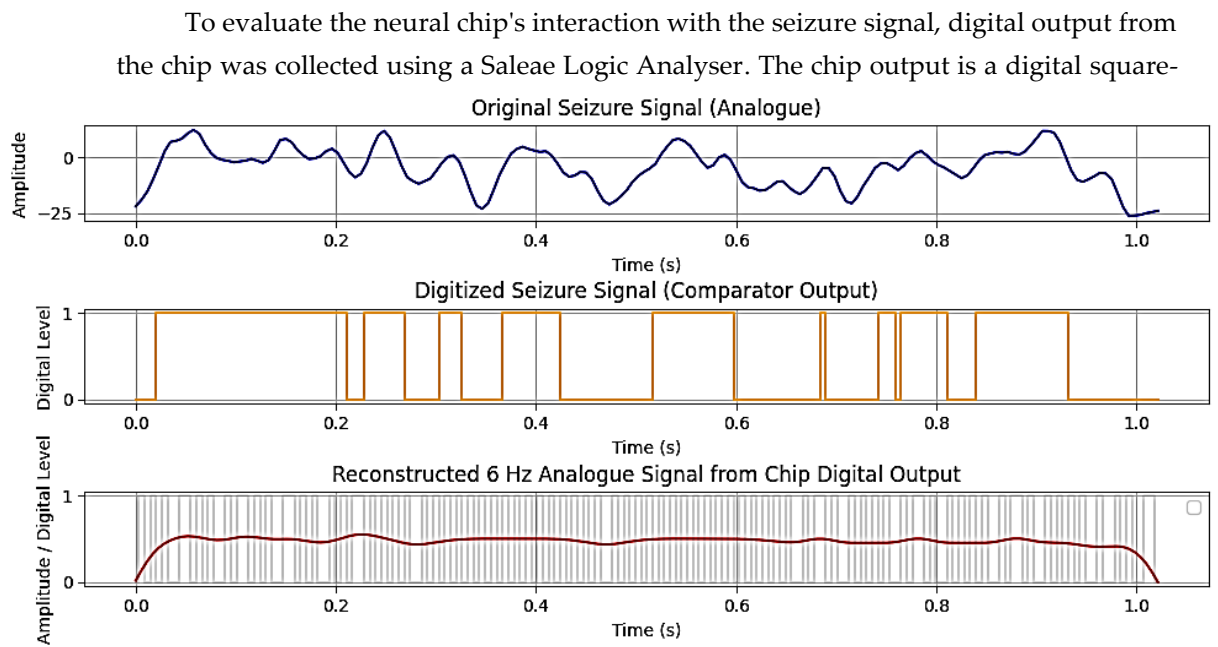

wave–like pattern intended to operate at 6 Hz, and the relevant information is stored in Channel 4 of the CSV file (as shown in Figure 12). Since the neural chip operates in the digital domain, the seizure signal is converted into digital form. A thresholding operation was used to convert the analogue seizure waveform into a binary representation. The threshold chosen was the mean value of the waveform. Any EEG amplitude greater than this threshold was assigned a digital high, and any value below was assigned a digital low. To demonstrate how the chip's 6 Hz rhythm could entrain the seizure activity, the resampled chip digital output was passed through a smoothing filter. This low-pass filtering converts the square-wave-like digital output into a smooth sinusoid-like analogue signal, as shown in Figure 13.

Figure 13: top (seizure signal), middle (digitized waveform), bottom (superimposed entrained signals)

The top plot shows the original averaged seizure signal as a continuous waveform, representing the typical shape of seizure activity. The middle plot shows the digitized version of that waveform after thresholding, displaying how the analogue EEG becomes a binary signal when fed into digital hardware. The bottom plot shows two superimposed signals: the actual digital pulses generated from the custom-designed neural chip, overlaid by the reconstructed analogue 6 Hz waveform. To further validate the entrainment principle, a frequency-domain analysis was computed using the Fourier transform, as shown in Figure 14. The spectrum of the original seizure waveform shows energy spread across multiple frequencies, which is characteristic of seizure activity. The spectrum of the reconstructed chip-derived analogue waveform, however, shows a clear peak at around 6 Hz.

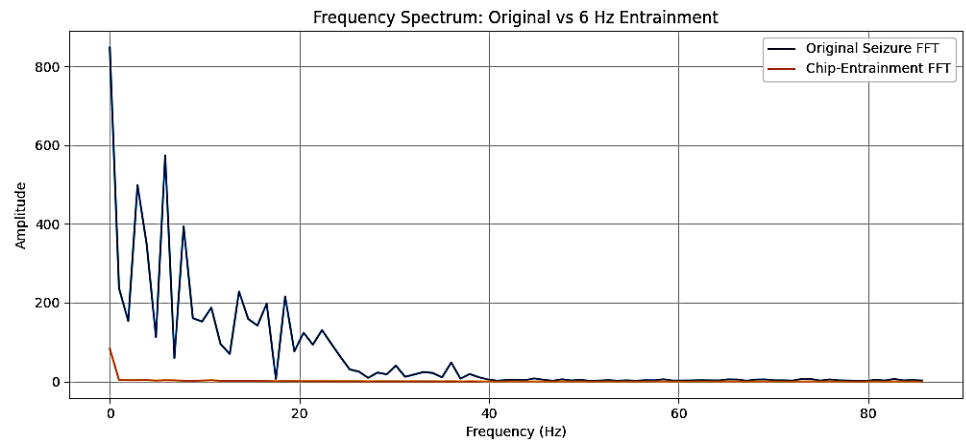

Figure 14: Frequency spectrum of the original seizure and entrained signal

## 3. IoT Connectivity with Mobile App

To further enhance the functionality of the HiLTS© platform, a mobile interface was developed using the Blynk IoT framework. The mobile application communicates wirelessly with an Arduino Wi-Fi module, enabling remote control of the neural chip's pulse generation, as shown in Figure 15. Upon detection and trigger of an epileptic seizure, the clinician can manually trigger the chip entrainment signal through the mobile application. The mobile interface also allows selection of specific output frequencies for the entrainment pulses, facilitating customization for different patient or research requirements. Real-time monitoring of the chip state is provided through the app, allowing confirmation of signal generation and trigger timing. This IoT-enabled integration demonstrates the potential for wireless, portable neuro-modulation, allowing user-controlled intervention in real-time brain rhythm regulation.

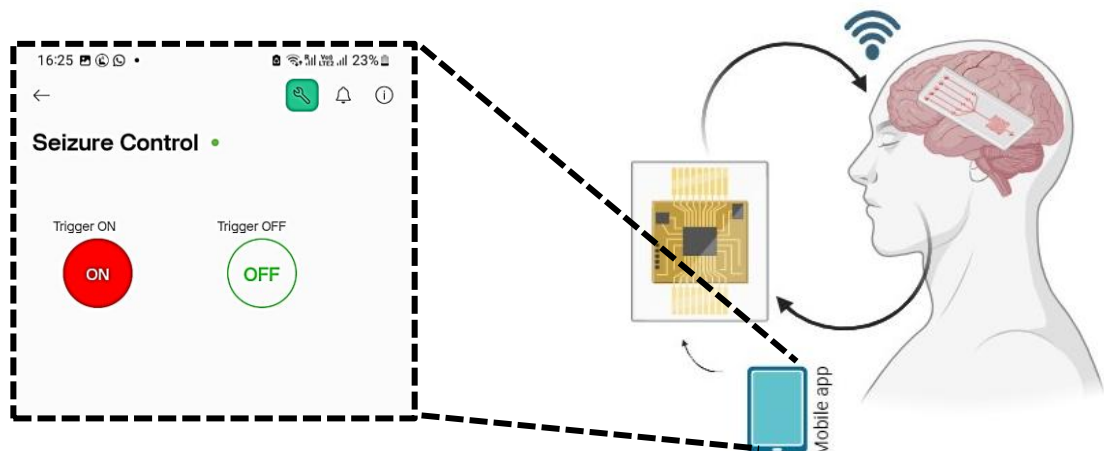

Figure 15: An overview of the mobile phone-based app control for chip entrainment signals

## 4. Discussion

The HiLTS© platform presented in this work establishes a foundational architecture for digital neuromodulation and precision medicine, demonstrating that entrainment of pathological brain rhythms can be achieved using a minimalist, low-power, custom-designed digital chip [34-37]. Unlike existing neuromodulation approaches that rely on analogue circuitry, high-voltage stimulation hardware, or proprietary platforms, this study shows that a fully digital oscillator implemented using a compact custom ASIC can impose a stable, narrow-band 6 Hz rhythm capable of overriding broadband seizure activity. By digitizing publicly available EEG seizure waveforms, interfacing them directly with the chip, and reconstructing the pulse train into an analogue signal, the work provides one of the earliest demonstrations of digital-only entrainment validated in hardware rather than simulation alone. The HiLTS© platform demonstrates that meaningful neural entrainment is possible even without analogue front-ends or complex stimulation electronics. This proof-of-concept is particularly valuable for future wearable neuromodulation, where power consumption, silicon area, and system simplicity are critical design constraints. The current implementation integrates several distinct modules, including vagus nerve stimulation, multiphase back-pain stimulation, seizure detection and entrainment, an SNN-based speech classifier, and a PicoRV32 bare metal processor within a 4 mm² silicon die area, and a total power consumption of 0.625 W. These results collectively highlight the feasibility of creating compact, multi-functional architectures for precision therapeutic interventions.

However, several components remain beyond the scope of the present study. The current prototype is complemented with a mobile IoT interface for manual triggering and does not yet incorporate real-time EEG acquisition, autonomous seizure detection, or a fully closed-loop workflow. Live analogue EEG signals were not directly fed into the chip;

instead, digitized seizure data were used to emulate neural inputs. To enable complete closed-loop operation, future versions of the hardware will include integrated high-fidelity analogue-to-digital (ADC) and digital-to-analogue (DAC) conversion stages, allowing the chip to receive real-time physiological input and deliver analogue entrainment signals without external reconstruction processes. Such mixed-signal extensions are essential for bridging biological interfaces with the digital neuromodulation core and will form a central part of subsequent development. Further work will focus on fabricating and characterizing the full multi-module system, expanding the entrainment engine to support multiple frequencies, integrating lightweight machine-learning models for autonomous detection and triggering, and performing comprehensive hardware characterization under real biological or neurophysiological conditions. Clinical validation, which is necessary for evaluating long-term safety and efficacy, remains an important future direction but lies outside the scope of the present paper.

Beyond its technical contributions, the HiLTS© platform represents one of the first neurotechnology systems of its kind to be developed in the United Arab Emirates, and to the best of the authors' knowledge, no comparable open or academic platform internationally offers such a unified, modular, and extensible architecture for precision neurotechnology research. Its design philosophy emphasizes openness, reproducibility, and accessibility, making it a valuable foundation for training new researchers and enabling future work on wearable diagnostic and therapeutic devices. The framework and methods described herein are currently under consideration for patent protection, further underscoring their novelty and translational potential. The proposed digital entrainment using custom ASIC hardware is a viable and promising approach to interrupting pathological neural rhythms. Although the present study focuses primarily on validating the digital entrainment capability of a single-frequency oscillator, the HiLTS© system platform lays the groundwork for future neuromodulation platforms that merge open-source hardware, mixed-signal interfaces, IoT connectivity, and real-time neurophysiological feedback into a unified and clinically relevant therapeutic system.

## 5. Summary

This work presents a novel digital neuromodulation platform that can be used to entrain pathological seizure activity through narrow-band rhythmic stimulation. By combining EEG signal analysis, digital ASIC design, hardware synthesis and simulation validation, and IoT-enabled control, the study establishes the framework for an end-to-end pipeline that bridges open-source design methodologies with real hardware implementation for healthcare engineering. The successful generation of stable 6 Hz entrainment rhythms and their interaction with digitized seizure signals provides strong evidence that digital-only stimulation architectures can influence seizure-like dynamics without the need for complex analogue circuitry. The current prototype serves as a foundational step toward a full precision-medicine neuromodulation system. Although the present work demonstrates entrainment using pre-recorded EEG and manual IoT-based triggering, it does not yet implement continuous real-time EEG acquisition, closed-loop seizure detection, or multi-frequency stimulation. Future iterations of the system will incorporate integrated ADC and DAC modules, expand the entrainment core to support multiple therapeutic bands, and complete the fabrication and characterization of the multi-module HiLTS© system-on-chip. These extensions are essential for enabling autonomous, real-time neurostimulation and for establishing the platform as a viable candidate for wearable, implantable and point-of-care diagnostic applications. Beyond its technical contributions, the HiLTS© platform is the first open, modular, and extensible neuromodulation architecture of its kind in the region, capable of supporting multi-diagnostic and multi-therapeutic research in a unified hardware platform. Rather than restricting access to the technology, which in the authors' opinion has often been the case in hardware development, its openness and modularity make it well suited for training new researchers and for accelerating innovation in digital neurotechnology and precision medicine. The methods and architecture presented here are the subject of a forthcoming patent application, reflecting the novelty and translational potential of the system.

**Funding:** This research was partly funded by the AURAK Seed Research Grant, Project Reference No. ENGR/007/26.

**Data Availability Statement:** No new data were created in this research; however, the author of the paper would be happy to share further information upon request with interested researchers.

**Acknowledgements:** The author acknowledges the use of Grammarly for Microsoft Office 6.8.263 and ChatGPT-4o mini in the process of translating and improving the clarity and quality of the English language in this manuscript. All ideas, including chip design, chip characterization, hardware simulation and synthesis, use of open-source tools, app development, and system-level integration and testing, are the intellectual property of the author of this paper. The author list is provisional and could be updated.

**Conflicts of Interest:** The author declares that at the time the paper was published, they had no known competing financial interests or personal relationships that could have appeared to influence the work reported in this paper. All images and diagrams are indirectly linked to a related patent and are copyrighted material and intellectual property of the author of this paper. To reproduce ideas, images and diagrams based on this work, written permission is required from the author of the paper.